\documentclass[10pt,final,journal]{IEEEtran}

\IEEEoverridecommandlockouts

\usepackage{cite}
\usepackage{amsmath,amssymb,amsfonts,newtxmath}
\usepackage{graphicx}
\usepackage{textcomp,booktabs,multirow}
\usepackage[caption=false,font=footnotesize]{subfig}

\newcommand{\expval}[1]{\operatorname{E}[#1]}

\begin{document}
	
\title{Quantum Two-Mode Squeezing Radar and Noise Radar: Correlation Coefficients for Target Detection}

\author{
	\IEEEauthorblockN{David Luong,~\IEEEmembership{Student Member, IEEE}, Sreeraman Rajan,~\IEEEmembership{Senior Member, IEEE}, and \\ Bhashyam Balaji,~\IEEEmembership{Senior Member, IEEE}}
	
	\thanks{D.\ Luong is with Defence Research and Development Canada, Ottawa, ON, Canada K2K 2Y7, and also with Carleton University, Ottawa, ON, Canada K1S 5B6. Email: david.luong3@carleton.ca.}
	\thanks{S.\ Rajan is with Carleton University, Ottawa, ON, Canada K1S 5B6. Email: sreeraman.rajan@carleton.ca.}
	\thanks{B.\ Balaji is with Defence Research and Development Canada, Ottawa, ON, Canada K2K 2Y7. Email: bhashyam.balaji@drdc-rddc.gc.ca.}
}

\maketitle

\begin{abstract}
	Quantum two-mode squeezing (QTMS) radars and noise radars detect targets by correlating the received signal with an internally stored recording. A covariance matrix can be calculated between the two which, in theory, is a function of a single correlation coefficient. This coefficient can be used to decide whether a target is present or absent. We can estimate the correlation coefficient by minimizing the Frobenius norm between the sample covariance matrix and the theoretically expected form of the matrix. Using simulated data, we show that the estimates follow a Rice distribution whose parameters are simple functions of the underlying, ``true'' correlation coefficient as well as the number of integrated samples. We obtain an explicit expression for the receiver operating characteristic curve that results when the correlation coefficient is used for target detection. This is an important first step toward performance prediction for QTMS radars.
\end{abstract}


\section{Introduction}

Quantum radars promise increased detection performance by exploiting phenomena unique to quantum physics. For this reason, they have been attracting significant attention over the past few years. Quantum illumination radars \cite{lloyd2008qi,tan2008quantum,england2018quantum,balaji2018qi}, which are based on a phenomenon called \emph{entanglement}, are a particularly well-studied class of quantum radars. An experimental implementation of a \emph{quantum two-mode squeezing} (QTMS) radar, a variant of quantum illumination radar, was recently demonstrated \cite{chang2018quantum,luong2019quantum,luong2019roc}. Although it is not a full quantum radar (it uses amplifiers which break the entanglement), it demonstrates all the necessary ingredients of an entanglement-based radar, including the generation of an entangled signal and the transmission of microwaves through free space. It constitutes strong evidence that practical quantum radars can be built.

QTMS radars are very similar to standard noise radars (e.g.\ \cite{dawood2001roc}) in that they rely on the correlation between two noise signals for target detection. In fact, the name adopted for QTMS radar in \cite{chang2018quantum} was \emph{quantum-enhanced noise radar}. One of the noise signals is transmitted toward a target, while the other is retained within the radar system. The latter is then correlated with the received signal. When the correlation is high, the target is declared to be present. 

The Pearson correlation coefficient, often known simply as \emph{the} correlation coefficient, is a natural measure of this correlation. In this paper, we investigate one method of extracting this coefficient and using it for target detection. Using simulated data, we obtain empirical analytical expressions for the probability distribution of the correlation coefficient. These expressions can be used to obtain an expression for the receiver operating characteristic (ROC) curve which would be obtained when the correlation coefficient is used as a detector function to distinguish whether a target is present or absent. Finally, we show that these expressions fit the experimental data obtained from the QTMS radar experiment described in \cite{luong2019roc}.

\section{Basic Operation of QTMS and Noise Radars}
\label{sec:summary}

\begin{figure}[t]
	\centerline{\includegraphics[width=.9\columnwidth]{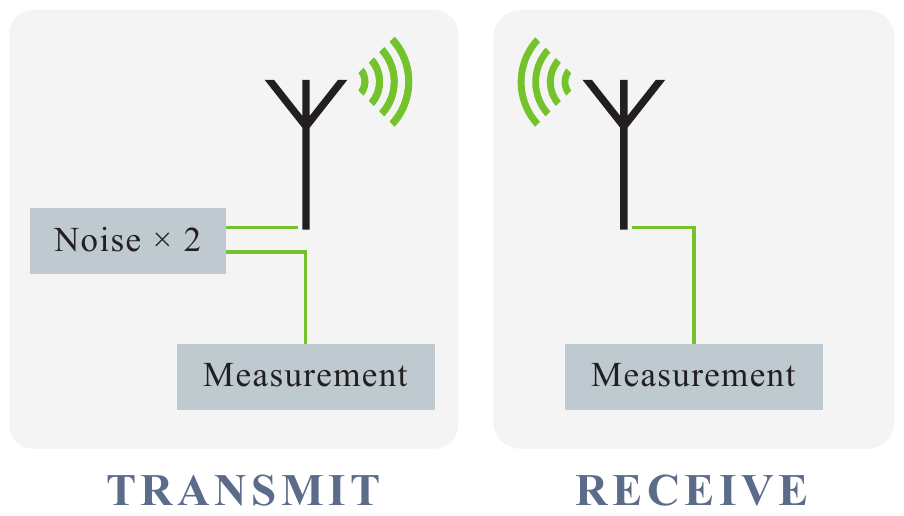}}
	\caption{Block diagram illustrating the basic idea of both QTMS radar and standard noise radar.}
	\label{fig:block_diagram_abstract}
\end{figure}

A simplified block diagram which applies to both quantum two-mode squeezing radar and noise radar is shown in Fig.\ \ref{fig:block_diagram_abstract}. For more details, see \cite{luong2019roc}.

Both radars follow this basic protocol:
\begin{enumerate}
	\item Produce two highly correlated noise signals.
	\item Record the time series of in-phase ($I$) and quadrature ($Q$) voltages for one of the signals and send it to the receiver. Transmit the other signal toward a target.
	\item Receive a signal and measure its $I$ and $Q$ voltages.
	\item Calculate a scalar which characterizes the correlation between the retained and received $I$ and $Q$ voltages. If this scalar exceeds a given threshold, declare the target to be present.
\end{enumerate}

The difference between QTMS radar and standard noise radar lies in step 1. A noise radar produces the two noise signals by splitting a single noise signal, whereas a QTMS radar produces a pair of entangled signals in a state known as \emph{two-mode squeezed vacuum}, in which the two signals can be thought of as highly correlated Gaussian noise. Surprisingly, quantum physics tells us that splitting a single signal does not result in a pair of perfectly correlated signals. Even ideal beamsplitters introduce an irreducible quantity of noise, known as \emph{quantum noise}. Entanglement causes the quantum noise between the two signals to be correlated, which allows us to achieve higher correlations. For more details on entanglement and quantum noise, see \cite{luong2019roc}.

In this paper, we focus on step 4, which is common to both QTMS radar and noise radar. Let us denote by $I_1$, $Q_1$ the in-phase and quadrature voltages for the received signal and $I_2$, $Q_2$ the voltages of the measurement record. We can then form the vector $x = [I_1, Q_1, I_2, Q_2]^T$ and the $4\times4$ covariance matrix $\expval{xx^T}$. It is from this covariance matrix that we will extract a correlation coefficient that can be used for target detection.

Note that $\expval{xx^T}$ is not related to the matrices that arise in array processing. In this paper, we consider only SISO (single input, single output) radars. The voltages that form the vector $x$ are raw, not filtered; indeed, $I_2$ and $Q_2$ can be thought of as a reference signal for the matched filtering of $I_1$ and $Q_1$.

\section{Covariance Matrix Structure for QTMS and Noise Radar Signals}
\label{sec:cov_mats}

It can be shown that, if all unwanted noise in a QTMS radar is additive white Gaussian noise (AWGN), the covariance matrix $\expval{xx^T}$ between the recorded and received QTMS radar signals has the form
\begin{equation} \label{eq:QTMS_cov}
	\begin{split}
		&R_\text{QTMS}(\sigma_1, \sigma_2, \rho, \phi) = \\
		&\begin{bmatrix}
			\sigma_1^2 & 0 & \rho \sigma_1 \sigma_2 \cos\phi & \rho \sigma_1 \sigma_2 \sin\phi \\
			0 & \sigma_1^2 & \rho \sigma_1 \sigma_2 \sin\phi & -\rho \sigma_1 \sigma_2 \cos\phi \\
			\rho \sigma_1 \sigma_2 \cos\phi & \rho \sigma_1 \sigma_2 \sin\phi & \sigma_2^2 & 0 \\
			\rho \sigma_1 \sigma_2 \sin\phi & -\rho \sigma_1 \sigma_2 \cos\phi & 0 & \sigma_2^2
		\end{bmatrix}
	\end{split}
\end{equation}
where $\sigma_1^2$ and $\sigma_2^2$ denote the measured signal powers for $I_1$ and $I_2$, $\phi$ is the phase between the received and recorded signals, and $\rho$ is what the Pearson correlation coefficient between $I_1$ and $I_2$ would be if $\phi = 0$ \cite{luong2019cov}. 

Given a general pair of signals, the cross-correlations between them would be characterized by four different correlation coefficients: $\rho(I_1,I_2)$, $\rho(I_1,Q_2)$, $\rho(Q_1,I_2)$, and $\rho(Q_1,Q_2)$. For QTMS radar signals, all of the cross-correlations are characterized by the two parameters $\rho$ and $\phi$. Of these, it is $\rho$---the magnitude of the correlation between the received and retained signals---which is most relevant to target detection. When a target is absent, the received signal is purely background noise, totally uncorrelated with the recorded signal. Therefore we expect $\rho = 0$ when a target is absent and $\rho \neq 0$ when a target is present. In the following sections of this paper, we will explore one way of extracting $\rho$ from measurement data and using it for target detection.

A similar covariance matrix can be calculated for a standard noise radar. This was done in \cite{dawood2001roc}, although their notation was different and their results were not expressed in the form of a matrix. Translated into the notation used in this paper, their result is
\begin{align} \label{eq:noiseradar_cov}
	\begin{split}
		&R_\text{noise}(\sigma_1, \sigma_2, \rho, \phi) = \\
		&\begin{bmatrix}
			\sigma_1^2 & 0 & \rho \sigma_1 \sigma_2 \cos\phi & \rho \sigma_1 \sigma_2 \sin\phi \\
			0 & \sigma_1^2 & -\rho \sigma_1 \sigma_2 \sin\phi & \rho \sigma_1 \sigma_2 \cos\phi \\
			\rho \sigma_1 \sigma_2 \cos\phi & -\rho \sigma_1 \sigma_2 \sin\phi & \sigma_2^2 & 0 \\
			\rho \sigma_1 \sigma_2 \sin\phi & \rho \sigma_1 \sigma_2 \cos\phi & 0 & \sigma_2^2
		\end{bmatrix} \!\! .
	\end{split}
\end{align}
The parameters are the same as for \eqref{eq:QTMS_cov}. The only difference between the two matrices is a change in the locations of the negative signs. We can recover \eqref{eq:QTMS_cov} from this matrix by taking the negative of all $Q_1$ voltages: $Q_1 \rightarrow -Q_1$. This does not affect the possible utility of $\rho$ for target detection. Thus, for the purposes of this paper, the difference between \eqref{eq:QTMS_cov} and \eqref{eq:noiseradar_cov} is insignificant. We will phrase our results in terms of QTMS radar because that was our primary motivation, but any result obtained for one can be readily applied to the other.

It is common to combine the real-valued voltages considered here into complex-valued voltages of the form $I_1 + j Q_1$, resulting in complex $2 \times 2$ covariance matrices. However, in the complex voltage representation, the off-diagonal elements of the QTMS covariance matrix equal zero; $\rho$ and $\phi$ drop out of the expression entirely \cite{luong2019cov}. We therefore prefer the real-valued representation in this paper.

\section{Estimation of the Correlation Coefficient}
\label{sec:estimation}

The most straightforward method of estimating the covariance matrix $\expval{xx^T}$ is to calculate the sample covariance matrix. If $x_1, \dots, x_N$ are $N$ snapshots of the vector $x = [I_1, Q_1, I_2, Q_2]^T$, then the sample covariance matrix is
\begin{align} \label{eq:sample_cov}
	\hat{S} = \frac{1}{N} \sum_{n=1}^N x_n x_n^T.
\end{align}
(Note that, for both QTMS radar and noise radar, the in-phase and quadrature voltages are normally distributed with mean zero \cite{dawood2001roc,luong2019roc,luong2019cov}.) The problem is that $\hat{S}$ is not guaranteed to be of the form \eqref{eq:QTMS_cov} because \eqref{eq:sample_cov} only constrains the resulting matrix to be positive semidefinite. Therefore, no unique way to extract a single correlation coefficient $\rho$ from $\hat{S}$ exists. We may, however, use $\hat{S}$ as part of a larger scheme for estimating the four parameters that appear in \eqref{eq:QTMS_cov}, including $\rho$. In this paper, our approach is to perform the minimization
\begin{equation} \label{eq:minimization}
	\min_{\sigma_1, \sigma_2, \rho, \phi} \left\lVert R_\text{QTMS}(\sigma_1, \sigma_2, \rho, \phi) - \hat{S} \right\rVert_F
\end{equation}
subject to $0 \leq \sigma_1$, $0 \leq \sigma_2$, $0 \leq \rho \leq 1$, and $-\pi < \phi \leq \pi$. (The subscript $F$ denotes the Frobenius norm.) The condition on $\rho$ may seem strange since the Pearson correlation coefficient can generally be negative, but an inspection of \eqref{eq:QTMS_cov} shows that, in this case, the sign of $\rho$ can be absorbed into $\phi$. For the purpose of target detection, we need only distinguish between $\rho = 0$ and $\rho \neq 0$; the sign is immaterial.

It may be possible to obtain the estimated values $\hat{\sigma}_1$, $\hat{\sigma}_2$, $\hat{\rho}$, and $\hat{\phi}$ as an explicit function of the entries of $\hat{S}$, but in practice this is probably impractical. The results of this paper were obtained numerically, using Mathematica's \texttt{NMinimize} function.

\section{Simulation Results and Analytical Approximations}

For the simulation portion of this study, we followed a procedure which is mathematically equivalent to the following:
\begin{enumerate}
	\item For a particular choice of $\sigma_1$, $\sigma_2$, $\rho$, and $\phi$, generate $N$ normally distributed random vectors with mean zero and covariance matrix $R_\text{QTMS}(\sigma_1, \sigma_2, \rho, \phi)$.
	\item Form the sample covariance matrix \eqref{eq:sample_cov}.
	\item Perform the minimization in \eqref{eq:minimization} to obtain an estimated correlation coefficient $\hat{\rho}$.
\end{enumerate}
We say ``mathematically equivalent'' because, instead of simulating $N$ random vectors, we simulated a random matrix $X$ drawn from the Wishart distribution $W_4(R_\text{QTMS}, N)$. Its probability density function is
\begin{equation}
	f(X|V, N) = \frac{ \det(X)^{(N-p-1)/2} e^{-\operatorname{tr}(V^{-1} X)/2} }{ 2^{Np/2} \det(V)^{N/2} \Gamma_p\left(\frac {N}{2}\right) }
\end{equation}
where $V$ is a $p \times p$ positive definite matrix and $\Gamma_p$ is the multivariate gamma function. Having done this, the normalized matrix $X/N$ is equivalent to the result of performing steps 1 and 2. The use of the Wishart distribution is a computationally easier method of performing this procedure.

In our simulations, we repeated the above procedure 50\,000 times for various values of $\rho$ and $N$ and plotted histograms of $\hat{\rho}$ for each set of parameters. A subset of our simulation results are presented in Table \ref{table:results}, and a selection of the histograms is given in Fig.\ \ref{fig:histograms}. These particular values and histograms were obtained for $\sigma_1 = 0.5$, $\sigma_2 = 2$, and $\phi = 0$, but substantially similar results were obtained for $\sigma_1 = 1$, $\sigma_2 = 3$, and $\phi = \pi/4$. This is expected; we do not expect the distribution of $\hat{\rho}$ to depend on $\sigma_1$, $\sigma_2$, or $\phi$. In Table \ref{table:results}, the mean and standard deviation of the obtained estimates $\hat{\rho}$ are denoted by $\expval{\hat{\rho}}$ and $s[\hat{\rho}]$, respectively. The parameters $\alpha$ and $\beta$ are explained below.

\subsection{Analytical Approximation: The Rice Distribution}
\label{subsec:analytical_approx}

We have empirically found that all of our histograms are well-approximated by the Rice distribution when $N$ is greater than approximately 100. The probability density function (PDF) for the Rice distribution is
\begin{align} \label{eq:rice_PDF}
	f(x | \alpha, \beta) = \frac{x}{\beta^2} \exp \! \left( -\frac{x^2 + \alpha^2} {2\beta^2} \right) I_0 \! \left( \frac{x\alpha}{\beta^2} \right)
\end{align}
where $I_0$ is the modified Bessel function of the first kind with order zero (not to be confused with the in-phase voltages $I_1$ and $I_2$). The final two columns of Table \ref{table:results} are maximum likelihood estimates of the Rice distribution parameters $\hat{\alpha}$ and $\hat{\beta}$ for each histogram of $\hat{\rho}$.

\begin{table}[t]
	\centering
	\caption{Simulation Results}
	\label{table:results}
	\begin{tabular}{cccccc}
		\toprule
		$\rho$ & $N$ & $\expval{\hat{\rho}}$ & $s[\hat{\rho}]$ & $\hat{\alpha}$ & $\hat{\beta}$ \\
		\midrule
		& $10^4$ & 0.008855 & 0.004637 & 0.002150 & 0.006902 \\
		0 & $5 \times 10^4$ & 0.003964 & 0.002075 & 0.000694 & 0.003126 \\
		& $10^5$ & 0.002803 & 0.001469 & 0.000266 & 0.002230 \\
		\midrule
		& $10^4$ & 0.009982 & 0.005150 & 0.005180 & 0.007047 \\ 
		0.005 & $5 \times 10^4$ & 0.006130 & 0.002747 & 0.005017 & 0.003158 \\ 
		& $10^5$ & 0.005544 & 0.002091 & 0.005005 & 0.002242 \\ 
		\midrule
		& $10^4$ & 0.012798 & 0.005981 & 0.009939 & 0.007098 \\ 
		0.01 & $5 \times 10^4$ & 0.010510 & 0.003067 & 0.009995 & 0.003160 \\ 
		& $10^5$ & 0.010261 & 0.002211 & 0.010006 & 0.002241 \\ 
		\midrule
		& $10^4$ & 0.050512 & 0.007055 & 0.050006 & 0.007091 \\ 
		0.05 & $5 \times 10^4$ & 0.050094 & 0.003149 & 0.049995 & 0.003152 \\ 
		& $10^5$ & 0.050049 & 0.002228 & 0.050000 & 0.002229 \\ 
		\midrule
		& $10^4$ & 0.100216 & 0.006998 & 0.099970 & 0.007006 \\ 
		0.1 & $5 \times 10^4$ & 0.100057 & 0.003133 & 0.100008 & 0.003134 \\ 
		& $10^5$ & 0.100008 & 0.002214 & 0.099984 & 0.002214 \\ 
		\midrule
		& $10^4$ & 0.500039 & 0.005294 & 0.500011 & 0.005294 \\ 
		0.5 & $5 \times 10^4$ & 0.499989 & 0.002373 & 0.499983 & 0.002373 \\ 
		& $10^5$ & 0.500008 & 0.001669 & 0.500005 & 0.001669 \\ 
		\midrule
		& $10^4$ & 0.799989 & 0.002536 & 0.799985 & 0.002536 \\ 
		0.8 & $5 \times 10^4$ & 0.799996 & 0.001140 & 0.799995 & 0.001140 \\ 
		& $10^5$ & 0.800005 & 0.000805 & 0.800005 & 0.000805 \\ 
		\midrule
		& $10^4$ & 0.900005 & 0.001341 & 0.900004 & 0.001341 \\ 
		0.9 & $5 \times 10^4$ & 0.899998 & 0.000601 & 0.899998 & 0.000601 \\ 
		& $10^5$ & 0.899999 & 0.000425 & 0.899999 & 0.000425 \\ 
		\midrule
		& $10^4$ & 0.990001 & 0.000141 & 0.990001 & 0.000141 \\ 
		0.99 & $5 \times 10^4$ & 0.990000 & 0.000063 & 0.990000 & 0.000063 \\ 
		& $10^5$ & 0.990000 & 0.000044 & 0.990000 & 0.000044 \\
		\bottomrule
	\end{tabular}
\end{table}

\begin{figure*}[p]
	\centering
	\subfloat[]{\includegraphics[width=\columnwidth]{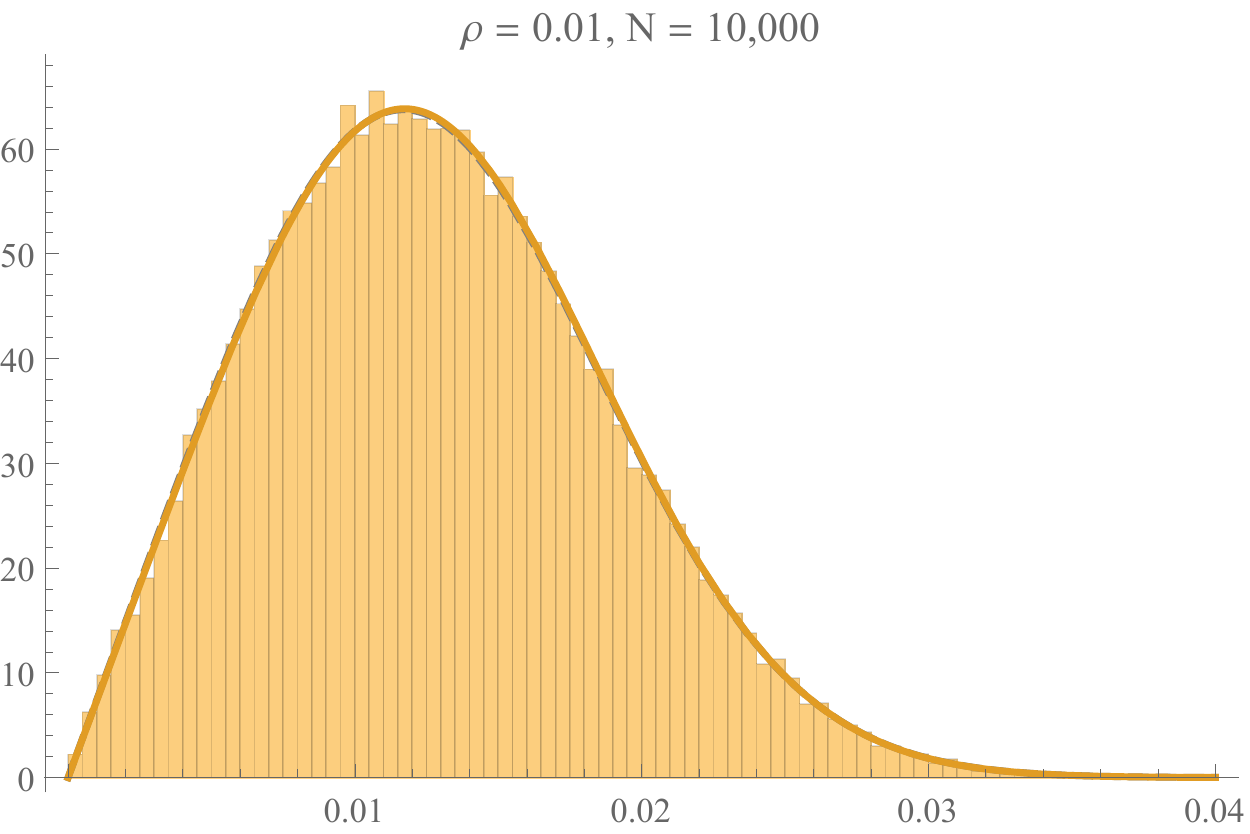}
		\label{subfig:histogram_001_10k}}
	\hfil
	\subfloat[]{\includegraphics[width=\columnwidth]{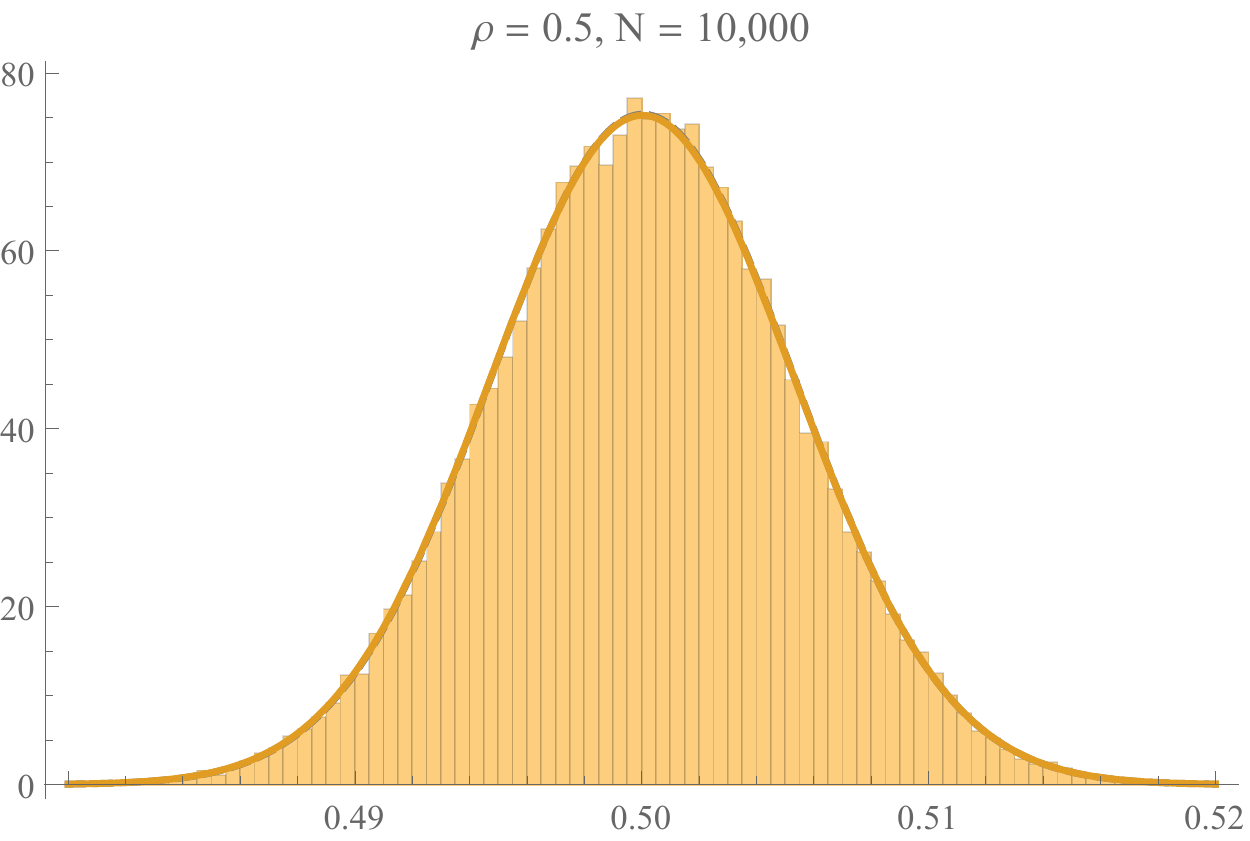}
		\label{subfig:histogram_05_10k}}
	\\
	\subfloat[]{\includegraphics[width=\columnwidth]{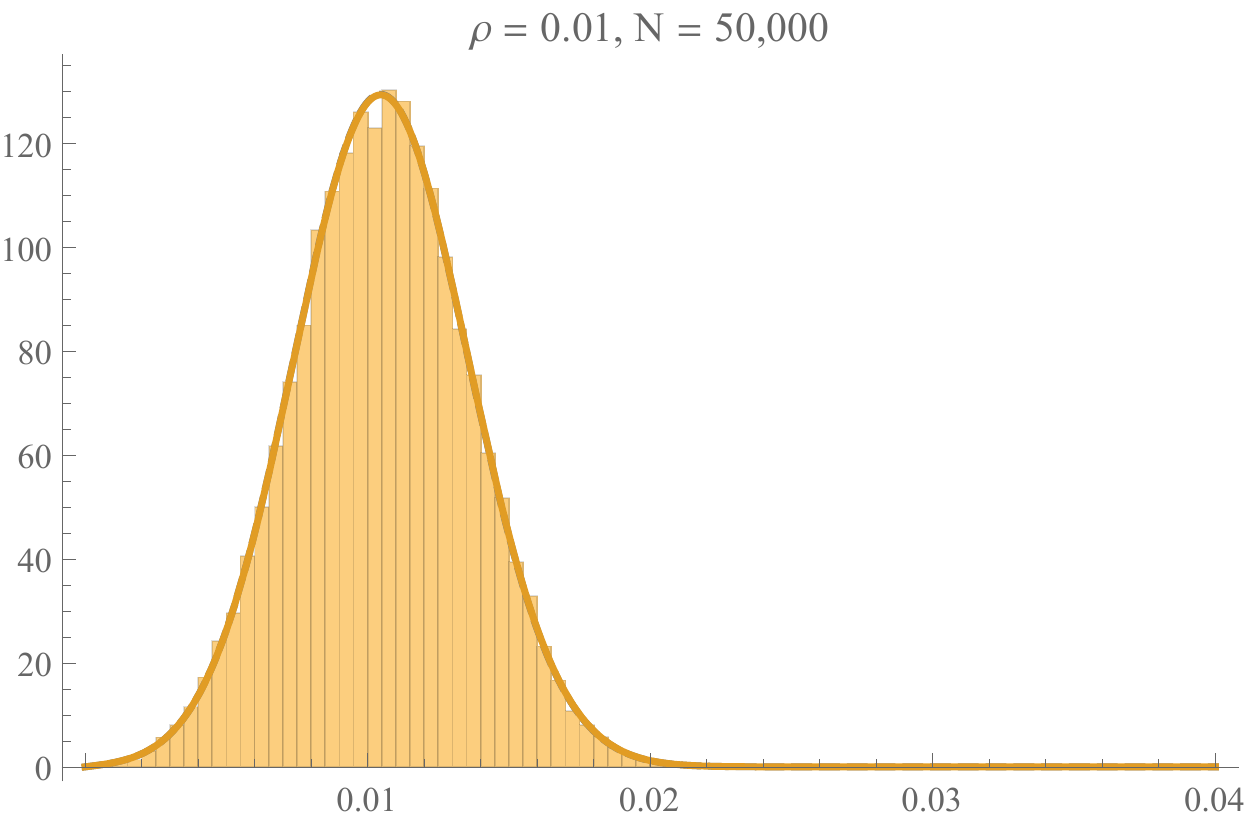}
		\label{subfig:histogram_001_50k}}
	\hfil
	\subfloat[]{\includegraphics[width=\columnwidth]{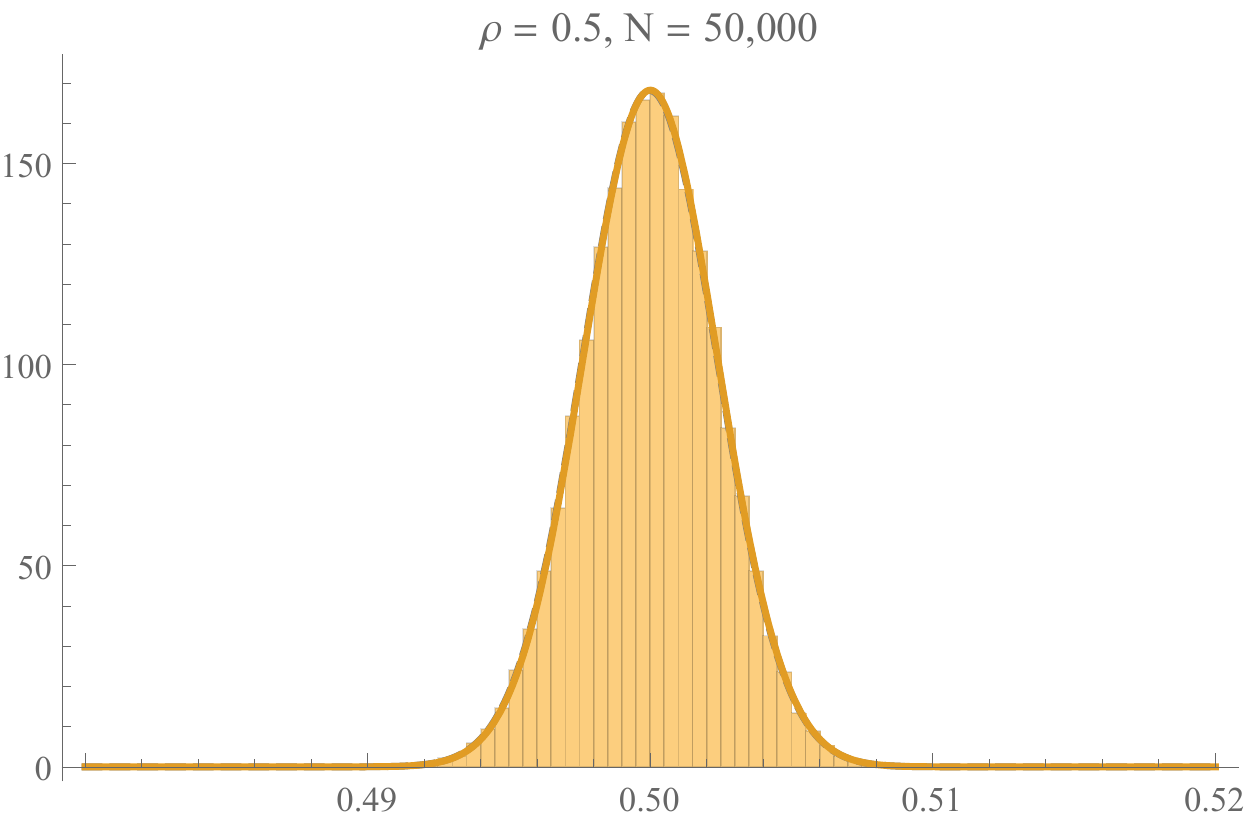}
		\label{subfig:histogram_05_50k}}
	\\
	\subfloat[]{\includegraphics[width=\columnwidth]{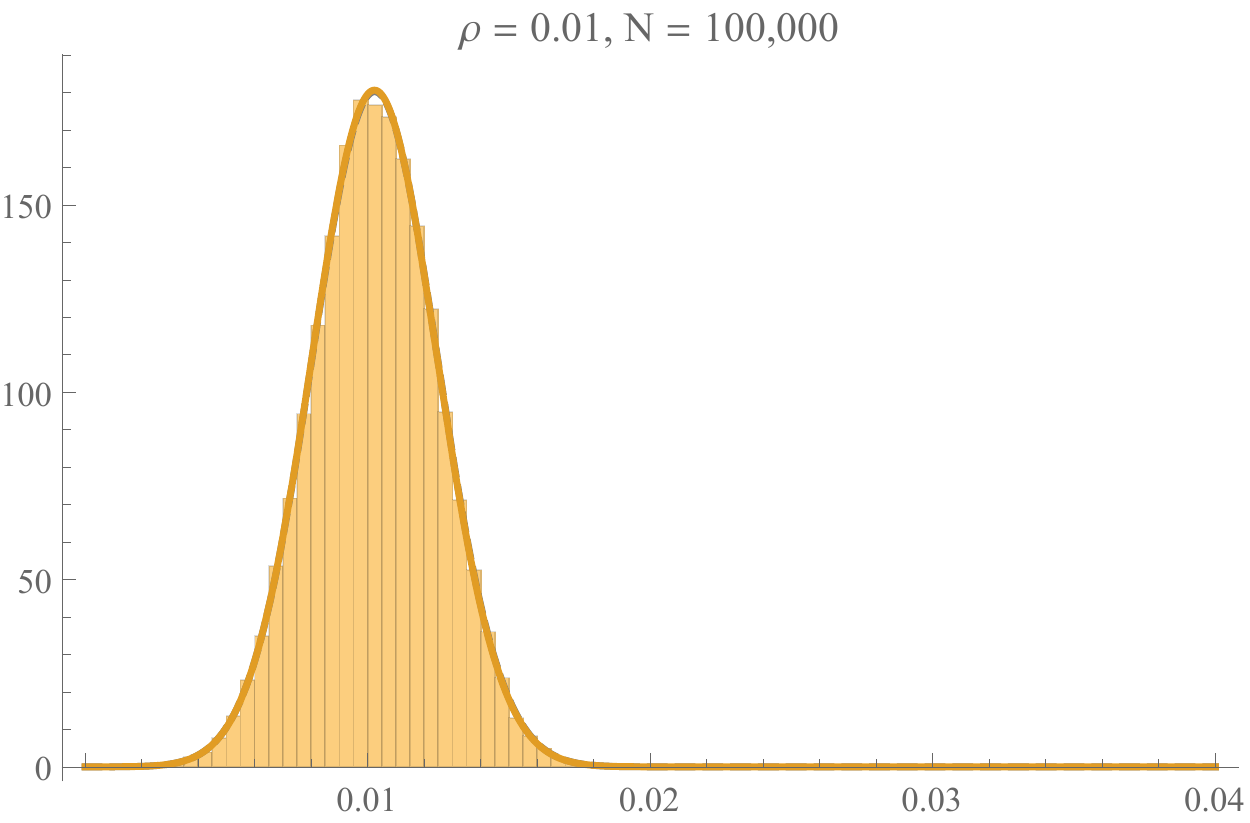}
		\label{subfig:histogram_001_100k}}
	\hfil
	\subfloat[]{\includegraphics[width=\columnwidth]{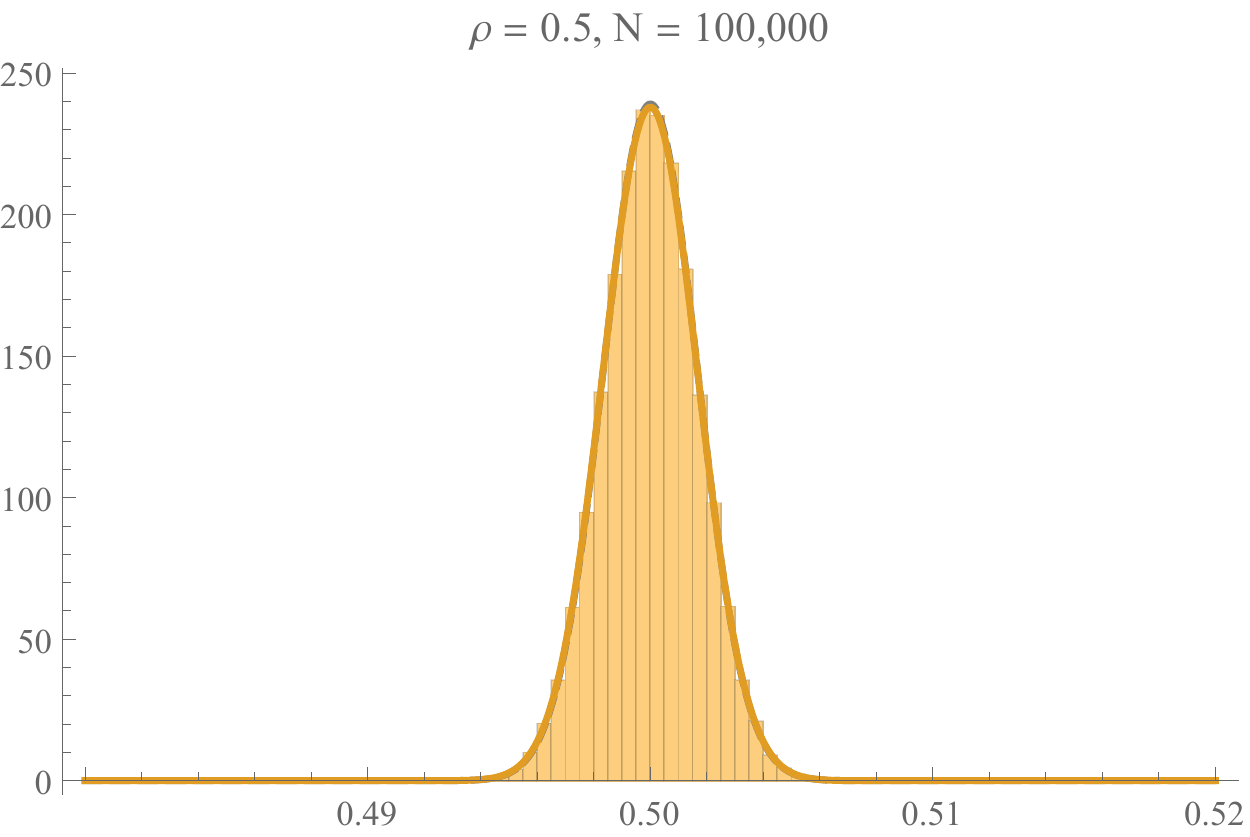}
		\label{subfig:histogram_05_100k}}
	\caption{Histograms of simulated $\hat{\rho}$ for varying values of $\rho$ and $N$. For all plots, $\sigma_1 = 0.5$, $\sigma_2 = 2$, and $\phi = 0$. Solid curves are Rice distribution PDFs with parameters given by \eqref{eq:analytic_params}. Dashed curves (only visible when zoomed in) are Rice distribution PDFs with parameters estimated directly from the histograms; they correspond almost exactly with the analytical approximations.}
	\label{fig:histograms}
\end{figure*}

Furthermore, we have found that the estimated values $\hat{\alpha}$ and $\hat{\beta}$ are well approximated by
\begin{subequations} \label{eq:analytic_params}
\begin{align}
	\alpha &= \rho \\
	\beta &= \frac{1-\rho^2}{\sqrt{2N}}.
\end{align}
\end{subequations}
When these values are inserted into the Rice distribution PDF \eqref{eq:rice_PDF}, the resulting plots give reasonably good fits to our simulated histograms, as seen in Fig.\ \ref{fig:histograms}.

There are two well-known probability distributions which are limiting cases of the Rice distribution, both of which are important in the context of target detection. When $\alpha/\beta$ is very large, it can be shown (via asymptotic expansion of the modified Bessel function $I_0$) that the Rice distribution approaches the normal distribution with mean $\alpha$ and standard deviation $\beta$. On the other hand, when $\alpha/\beta$ is very small, the Rice distribution approaches the Rayleigh distribution with scale parameter $\beta$. In fact, the Rayleigh distribution is a special case of the Rice distribution, obtained when $\alpha = 0$. This explains our previous finding in \cite{luong2019estimating} that the histograms for $\hat{\rho}$ appeared to be either Rayleigh distributed or normally distributed depending on the value of $\rho$.

The approximate analytical formulas in \eqref{eq:analytic_params} imply that 
\begin{equation}
	\frac{\alpha}{\beta} = \frac{\rho}{1-\rho^2} \sqrt{2N},
\end{equation}
which means that the Rice distribution approaches the normal distribution when $\rho$ or $N$ are very large. This behavior can be seen in Fig.\ \ref{fig:histograms}. Indirectly this can be seen in Table \ref{table:results} as well: as $\rho$ increases, the mean and standard deviation of $\hat{\rho}$ approaches $\hat{\alpha}$ and $\hat{\beta}$, respectively. This is expected because $\alpha$ and $\beta$ are the mean and standard deviation of the normal approximation. Conversely, for parameters where the normal approximation does not hold, $\expval{\hat{\rho}}$ and $s[\hat{\rho}]$ can be quite different from $\hat{\alpha}$ and $\hat{\beta}$; this too can be seen in Table \ref{table:results}.

It should be noted that the Rice distribution cannot represent the \emph{exact} distribution of $\hat{\rho}$ because the Rice distribution PDF is nonzero for all $x \geq 0$ whereas the definition of the Pearson correlation coefficient implies $\rho \leq 1$. However, this is not a grave concern in practice. When $\rho$ is large, the distribution becomes very narrow. This ensures that the portion of the Rice PDF where $x > 1$ is negligible. Similarly, when $\rho$ is small, the fact that the distribution is peaked near $x = 0$ again ensures that when $x > 1$, the Rice PDF is negligible. For a more rigorous argument, see Appendix \ref{sec:appendix_prob_greater_than_1}.

We may also note that the expressions \eqref{eq:rice_PDF} and \eqref{eq:analytic_params} have the merit of being relatively simple. Exact PDFs for quantities much like the estimated correlation coefficient $\hat{\rho}$ considered here are to be found in \cite{dawood2001roc} and \cite{gierull2004statistical}, but their expressions are more complicated. The preceding formulas may be compared, for example, with the distribution of the envelope detector output $Z$ described in \cite{dawood2001roc}:
\begin{equation}\label{eq:dawood_PDF}
	\begin{split}
		p(Z) &= \frac{2^{N+3} \tilde{Z}^N}{\sigma_1\sigma_2 (1-\rho^2) (N-1)!} K_{N-1}\! \left( \frac{4\tilde{Z}}{1-\rho^2} \right) \! I_0\! \left( \frac{4\rho\tilde{Z}}{1-\rho^2} \right)
	\end{split}
\end{equation}
where $\tilde{Z}$ is the normalized detector output $Z/(\sigma_1\sigma_2)$ and $K_{N-1}$ is the modified Bessel function of the second kind of order $N-1$; all other variables are as defined previously. Clearly, our Rice distribution approximation is more tractable. (We reiterate, however, that the results in \cite{dawood2001roc} and \cite{gierull2004statistical} were exact, whereas our result is an approximation.) In particular, we can use it to obtain closed-form expressions for receiver operating characteristic curves, as will be shown below.

\subsection{Target Detection; ROC Curves}

\begin{figure*}[t!]
	\centering
	\subfloat[]{\includegraphics[width=\columnwidth]{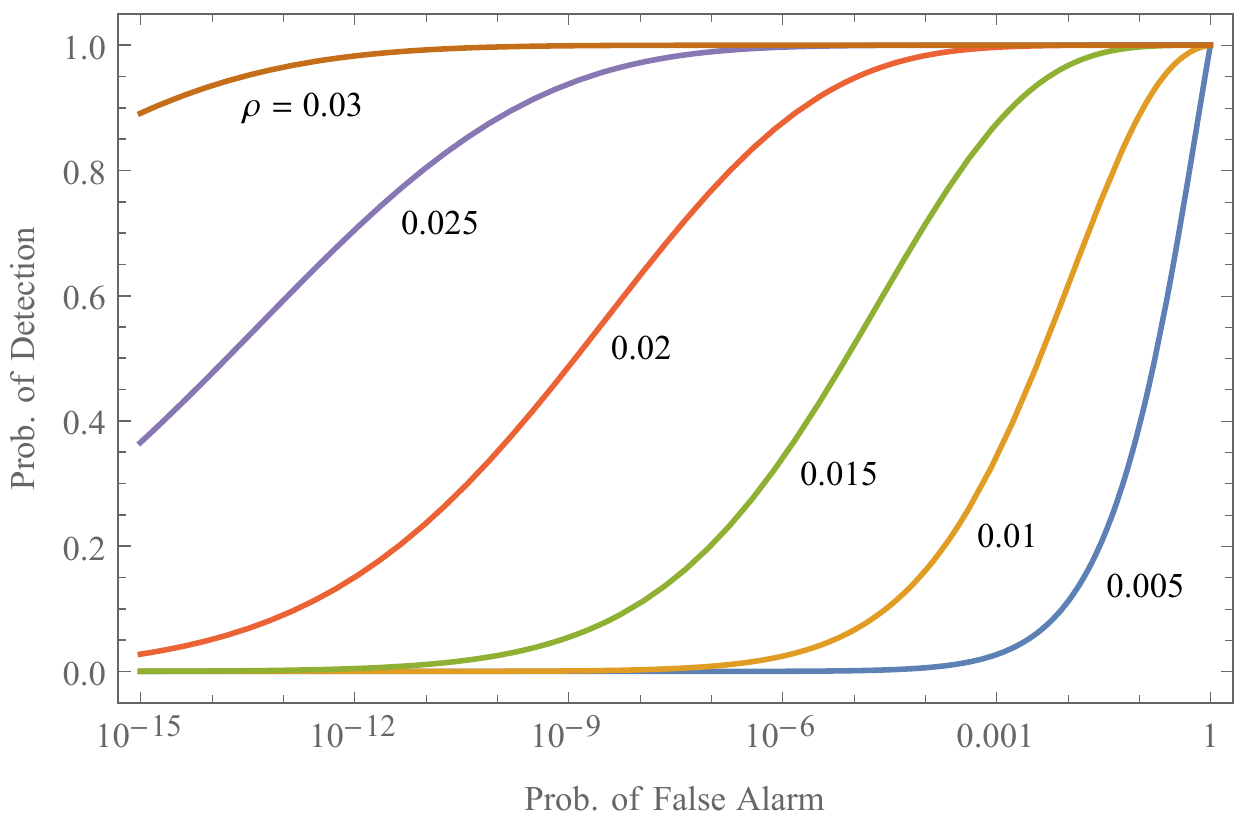}
		\label{subfig:ROC_sim_rho}}
	\hfil
	\subfloat[]{\includegraphics[width=\columnwidth]{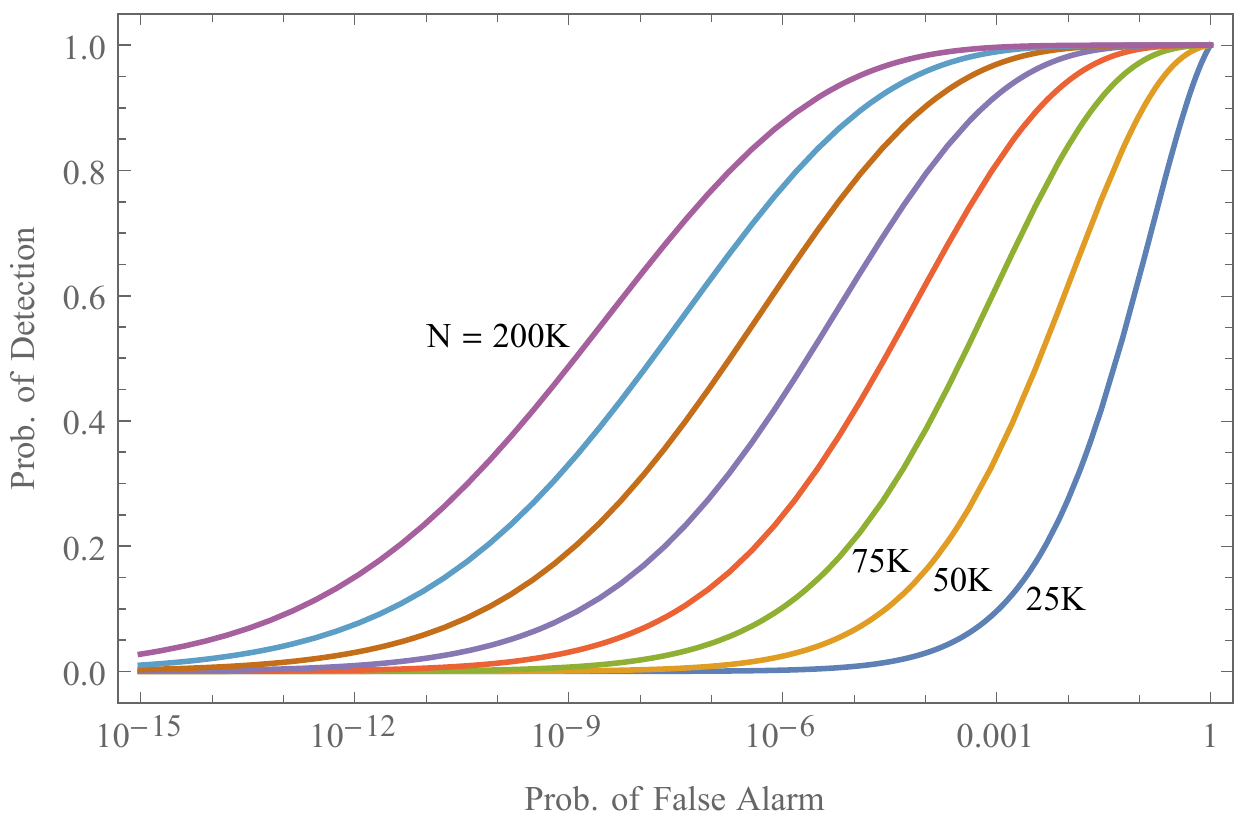}
		\label{subfig:ROC_sim_N}}
	\caption{Receiver operating characteristic curves when $\hat{\rho}$ is used as a detector function, calculated using \eqref{eq:ROC}. In (a), the curves from right to left correspond to $\rho$ = 0.005, 0.01, 0.015, 0.02, 0.025, and 0.03, respectively, while $N$ is fixed at 50\,000. In (b), the curves from right to left correspond to $N$ = 25\,000, 50\,000, 75\,000, 100\,000, 125\,000, 150\,000, 175\,000, and 200\,000, respectively, while $\rho$ is fixed at 0.01.}
	\label{fig:ROC_sim}
\end{figure*}

As mentioned earlier, we wish to use $\rho$ to distinguish whether a target is present or absent. These cases correspond to $\rho > 0$ and $\rho = 0$, respectively. In Sec.\ \ref{sec:estimation}, we have outlined how to obtain an estimate of $\rho$. Once we have such an estimate, we can decide between the two cases by setting some threshold $T$ and declaring a detection if $\hat{\rho} > T$.

Under the Rice approximation outlined above, we can calculate the probabilities of detection and of false alarm. The results can be combined to obtain a receiver operating characteristic (ROC) curve, which summarizes the performance of $\rho$ as a detector function. We begin by noting that the cumulative density function (CDF) of the Rice distribution is given by
\begin{equation}
	F(x | \alpha, \beta) = 1 - Q_1 \! \left( \frac{\alpha}{\beta}, \frac{x}{\beta} \right)
\end{equation}
where $Q_1$ denotes the Marcum $Q$-function. Substituting the values in \eqref{eq:analytic_params} gives
\begin{equation} \label{eq:rho_cdf}
	F(x | \rho, N) = 1 - Q_1 \! \left( \frac{\rho \sqrt{2N}}{1 - \rho^2}, \frac{x \sqrt{2N}}{1 - \rho^2} \right).
\end{equation}
When the target is absent, $\rho = 0$ and the Rice distribution CDF reduces to the Rayleigh distribution CDF, yielding
\begin{equation}
	F(x | 0, N) = 1 - e^{-N x^2}\! .
\end{equation}
For a given threshold $T$, the probability of false alarm is
\begin{equation}
	p_\text{FA}(T) = 1 - F(T | 0, N) = e^{-N T^2} \! ,
\end{equation}
which can easily be inverted to give
\begin{equation} \label{eq:threshold}
	T = \sqrt{- \frac{\ln p_\text{FA}}{N}}.
\end{equation}
The probability of detection, given a fixed value of $\rho$ and a threshold $T$, is
\begin{equation}
	p_\text{D}(T) = 1 - F(T | \rho, N).
\end{equation}
Upon substituting \eqref{eq:threshold}, we find that
\begin{equation} \label{eq:ROC}
	p_\text{D}(p_\text{FA} | \rho, N) = Q_1 \! \left( \frac{\rho \sqrt{2N}}{1 - \rho^2}, \frac{\sqrt{-2 \ln p_\text{FA}}}{1 - \rho^2} \right) \! .
\end{equation}
Remarkably, we have been able to obtain an explicit---and relatively simple---expression for $p_\text{D}$ as a function of $p_\text{FA}$.

Fig.\ \ref{fig:ROC_sim} shows ROC curves calculated using this formula. Plot (a) shows how the curve changes with $\rho$, while (b) shows how it changes with $N$. The two plots cannot be directly compared because $\rho$ and $N$ are fundamentally different quantities, but they do suggest that increasing the correlation of the signals results in more dramatic gains than increasing the integration time. This can be seen from \eqref{eq:analytic_params} as well: $\beta$ effectively controls the width of the Rice distribution, and it decreases with the square root of $N$ but quadratically with $\rho$. Of course, in practice certain tradeoffs need to be made between the two: increasing $\rho$ may require designing and building entirely new radar systems, while increasing $N$ indefinitely is impractical for many radar applications.

\section{Experimental Results}
\label{sec:experimental}

\begin{figure*}[t!]
	\centering
	\subfloat[]{\includegraphics[width=\columnwidth]{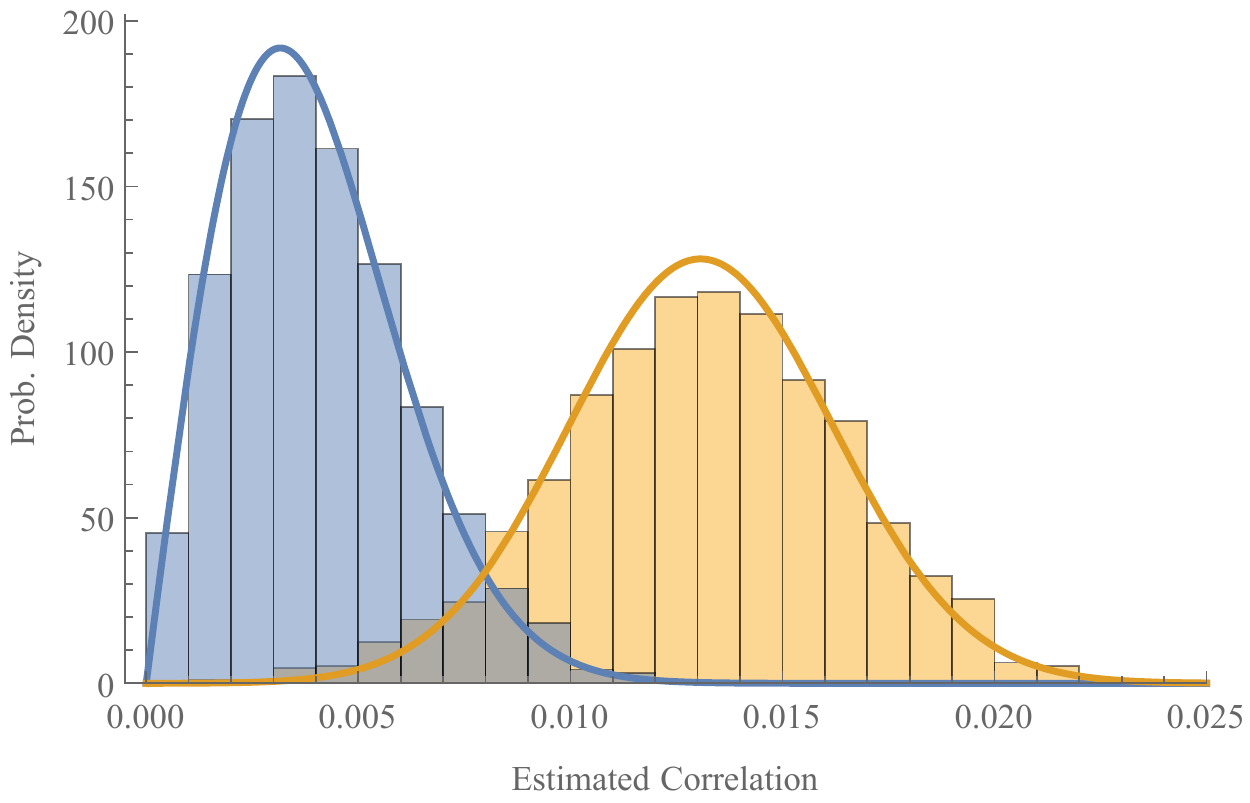}
		\label{subfig:histogram_wilson_q}}
	\hfil
	\subfloat[]{\includegraphics[width=\columnwidth]{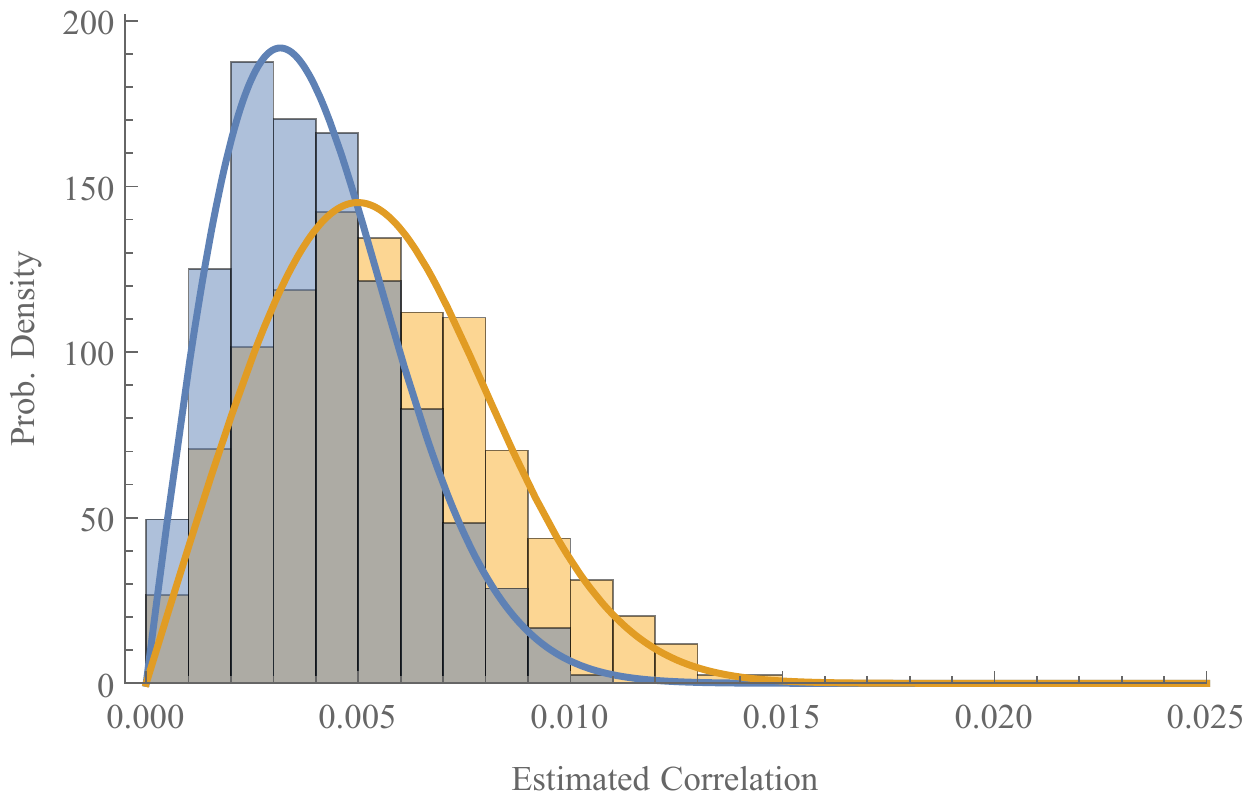}
		\label{subfig:histogram_wilson_c}}
	\caption{Histograms of $\hat{\rho}$ calculated from an experimental implementation of (a) a quantum two-mode squeezing radar and (b) a two-mode noise radar. Solid curves are Rice distribution PDFs with parameters given by \eqref{eq:analytic_params} with $N = 50\,000$. Blue histograms are for when the radars are off ($\rho = 0$). Orange histograms are for when the radars are on, in which case we have used $\rho_\text{QTMS} = 0.0127$, $\rho_\text{TMN} = 0.00419$.}
	\label{fig:histograms_wilson}
\end{figure*}

\begin{figure}[t]
	\centering
	\includegraphics[width=\columnwidth]{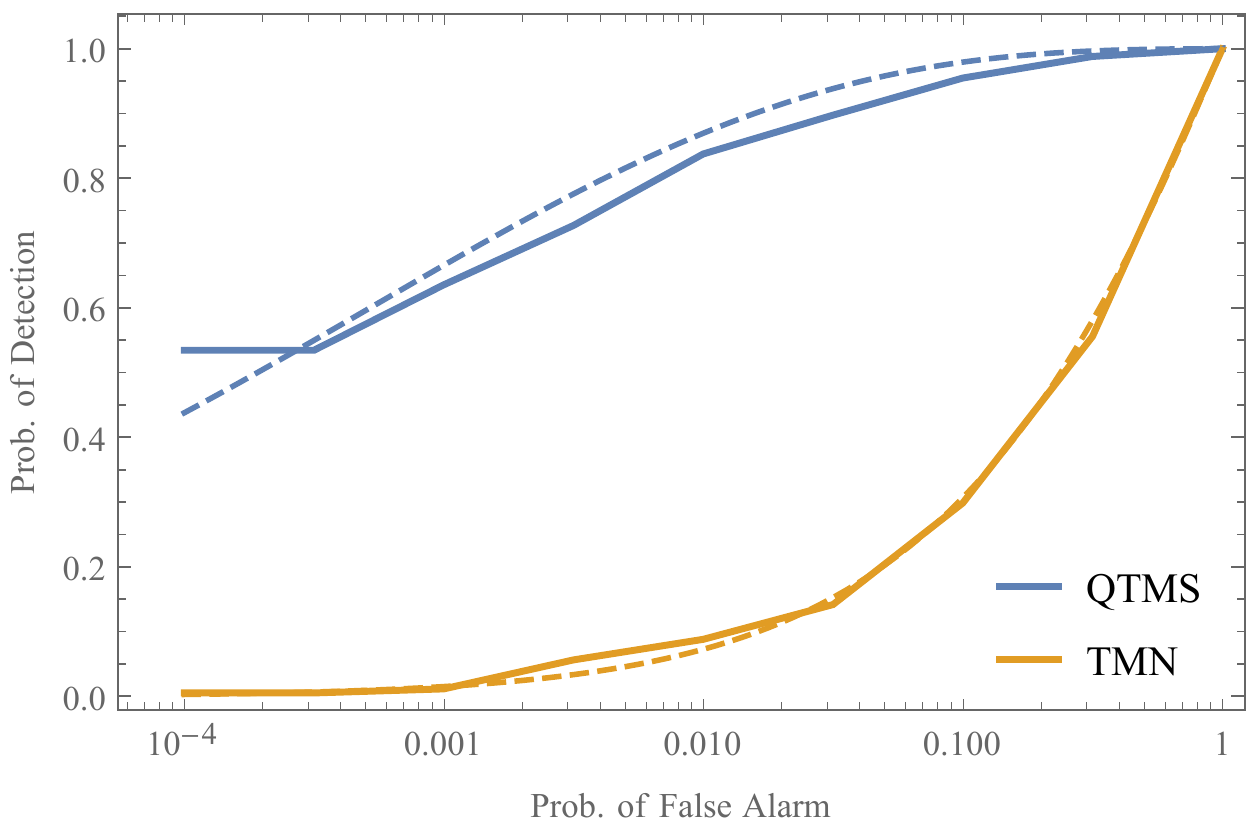}
	\caption{Receiver operator characteristic curves for an experimental implementation of a quantum two-mode squeezing radar (blue) and a two-mode noise radar (orange). In both cases, $N = 50\,000$. Solid curves were calculated directly from experimental data; dashed curves using \eqref{eq:ROC}.}
	\label{fig:ROC_wilson}
\end{figure}

In \cite{luong2019roc}, we described an experimental implementation of the QTMS radar protocol. It followed the basic idea described in Sec.\ \ref{sec:summary}, with the highly correlated noise signals being generated using a source of entangled microwaves. We compared the results to a radar which is similar in all respects except that the generated signal was \emph{classically} correlated without being entangled. (The pair of signals was generated by mixing Gaussian noise with a sinusoidal carrier and extracting the two sidebands.) The latter was similar to a standard noise radar except that the correlation structure is the same as for QTMS radar, which is given in \eqref{eq:QTMS_cov}. We call it a \emph{two-mode noise radar} (TMN radar). In both cases, the transmitted signal was sent through free space and received directly without being reflected off a target.

Here, we show that the experimental results obtained for the QTMS and TMN radars support our contention that the distribution of $\hat{\rho}$ is well-approximated by the Rice distribution with parameters given by \eqref{eq:analytic_params}. To this end, we applied the procedure in Sec.\ \ref{sec:estimation} to the experimental data obtained in \cite{luong2019roc} and produced four histograms of $\hat{\rho}$: one for the case where the QTMS radar was turned on, one for when it was turned off, and similarly for the TMN radar. In all four cases, we integrated $N = 50\,000$ samples when generating the sample covariance matrix \eqref{eq:sample_cov}. The results, with corresponding Rice distribution PDFs, are shown in Fig.\ \ref{fig:histograms_wilson}.

In our experiment, we did not know \emph{a priori} what the true value of $\rho$ was for the QTMS and TMN radars when they were turned on. Therefore, when comparing the corresponding histograms to the Rice approximation, it was necessary to obtain values for $\rho$ in order to substitute them into \eqref{eq:analytic_params}. We found via maximum-likelihood parameter estimation that, for our experimental data, $\rho_\text{QTMS} \approx 0.0127$ and $\rho_\text{TMN} \approx 0.00419$. The histograms in Fig.\ \ref{fig:histograms_wilson} indeed appear to be well-approximated by the Rice distribution when these values of $\rho$ were used. As for the histograms obtained when the radars were off, we may assume $\rho = 0$. In this case, no parameter estimation was necessary and we could plot the PDFs directly. In this case, too, the Rice distribution is a good approximation.

Fig.\ \ref{fig:ROC_wilson} shows ROC curves calculated directly from experimental data together with ROC curves calculated using \eqref{eq:ROC}. The agreement is quite good, and the deviations can be attributed to experimental imperfections together with statistical fluctuations resulting from the relatively small amount of experimental data that was available for this analysis.

\section{Conclusion}

In this paper, we explored the possibility of using certain correlation coefficients that arise in QTMS radar or noise radar for target detection. We saw that, in both types of radars, the covariance between the received and recorded signals can be described by matrices of the form \eqref{eq:QTMS_cov} or \eqref{eq:noiseradar_cov}. Both matrices are parameterized by a single correlation coefficient $\rho$ which characterizes the magnitude of the correlations between the received and recorded signals. Because this coefficient is expected to be nonzero when a target is present and zero when the target is absent, it is a reasonable detector function for target detection.

In order to detect a target, it is necessary to estimate $\rho$ from observation data. The method we use in this paper is to minimize the Frobenius norm between the structured covariance matrix \eqref{eq:QTMS_cov} and the sample covariance matrix directly calculated from the data, as prescribed in \eqref{eq:minimization}. We found from simulated data that the estimated values $\hat{\rho}$ are well-approximated by the Rice distribution. We were even able to find approximate analytical formulas for the parameters of the Rice distribution PDF as a function of the underlying, ``true'' correlation coefficient $\rho$ and the number of integrated samples $N$. This enabled us to obtain an analytic expression for the ROC curve that would be obtained when $\rho$ were used as a detector function. These results were confirmed using experimental data.

Although the theory developed in this paper bears a strong resemblance to envelope detection in conventional radars, the underlying physical quantities are quite different: there, the Rice distribution describes the amplitude of the received voltages, whereas here it describes the correlation of those voltages with a measurement record. It is intriguing, nevertheless, that both cases end up being described by the Rice distribution. It may even be possible to use this correspondence to extend existing results in radar theory to noise radars and quantum radars. This point deserves further scrutiny.

We believe that this work lays the foundation for a new way to understand any radar which operates according to the protocol summarized in Sec.\ \ref{sec:summary}. This includes standard noise radars, TMN radars, and QTMS radars. There are many directions in which our ideas can be extended. For example, we could attempt to combine this work with the radar range equation in order to obtain performance predictions for a QTMS radar. (This would require a knowledge of how $\rho$ is affected when the signal is reflected off a target.) Such a performance prediction framework would be of great utility in pinning down the precise regime in which quantum radars can offer a benefit over currently existing radars.

\appendices
\section{}
\label{sec:appendix_prob_greater_than_1}

In this appendix, we calculate an upper bound on the probability that $\hat{\rho} > 1$ under the Rice distribution approximation described in Sec.\ \ref{subsec:analytical_approx}.

We begin by noting that, according to \eqref{eq:rho_cdf}, the probability of obtaining a value greater than 1 is given by
\begin{equation} \label{eq:appendix_marcumq}
	P(x > 1) = Q_1 \! \left( \frac{\rho \sqrt{2N}}{1 - \rho^2}, \frac{\sqrt{2N}}{1 - \rho^2} \right) \! .
\end{equation}
An upper bound for the Marcum $Q$-function was given in \cite{simon2000marcumq}:
\begin{equation}
	Q_1(a, b) \leq \exp \! \left[ -\frac{(b - a)^2}{2} \right]
\end{equation}
whenever $0 \leq a < b$. This condition holds here because $0 \leq \rho \leq 1$, as stated in Sec.\ \ref{sec:estimation}, and \eqref{eq:appendix_marcumq} is undefined for $\rho = 1$. We therefore have
\begin{equation}
	P(x > 1) \leq e^{-N/(1 + \rho)^2} \leq e^{-N/4}.
\end{equation}
From this expression we find that, if $N > 100$ (the initial assumption made in Sec.\ \ref{subsec:analytical_approx}), the probability that $\hat{\rho} > 1$ under the Rice approximation is no greater than $e^{-25} \approx 1.39 \times 10^{-11}$. For all intents and purposes, therefore, the fact that the Rice distribution predicts a nonzero probability of obtaining values greater than 1 is not a major concern.

\section*{Acknowledgment}

We thank C.\ Wilson, C.\ W.\ S.\ Chang, and A.\ M.\ Vadiraj for providing us the experimental data which was used to obtain the results in Sec.\ \ref{sec:experimental}.

\bibliographystyle{ieeetran}
\bibliography{qradar_refs}
	
\end{document}